\documentclass[10pt,twocolumn,letterpaper]{article}

\usepackage{cvpr}
\usepackage{times}
\usepackage{epsfig}
\usepackage{graphicx}
\usepackage{amsmath}
\usepackage{amssymb, url, bm}

\usepackage[pagebackref=true,breaklinks=true,letterpaper=true,colorlinks,bookmarks=false]{hyperref}

\cvprfinalcopy 

\def\cvprPaperID{414} 

\ifcvprfinal\fi
\begin{document}

\title{Edge-Preserving Piecewise Linear Image Smoothing Using Piecewise Constant Filters}

\author{Wei Liu$^1$ ~ ~ ~
    Wei Xu$^1$ ~ ~  Xiaogang Chen$^{2}$ ~ ~  Xiaolin Huang $^1$ ~ ~  Chuanhua Shen$^{3}$ ~ ~ Jie Yang$^{1}$\\
$^1$Shanghai Jiao Tong University,  China
~ ~ ~
$^2$University of Shanghai for Science and Technology,  China \\
$^3$The University of Adelaide, Australia ~ ~ ~ ~ ~ ~
}

\maketitle

\begin{abstract}
Most image smoothing filters in the literature assume a piecewise constant model of smoothed output images. However, the piecewise constant model assumption can cause artifacts such as gradient reversals in applications such as image detail enhancement, HDR tone mapping, etc. In these applications, a piecewise linear model assumption is more preferred. In this paper, we propose a simple yet very effective framework to smooth images of piecewise linear model assumption using classical filters with the piecewise constant model assumption. Our method is capable of handling with gradient reversal artifacts caused by the piecewise constant model assumption. In addition, our method can further help accelerated methods, which need to quantize image intensity values into different bins, to achieve similar results that need a large number of bins using a much smaller number of bins. This can greatly reduce the computational cost. We apply our method to various classical filters with the piecewise constant model assumption. Experimental results of several applications show the effectiveness of the proposed method.
\end{abstract}

\vspace{-1em}
\section{Introduction}
\label{SecIntroduction}
\begin{figure}
  \includegraphics[width=1\linewidth]{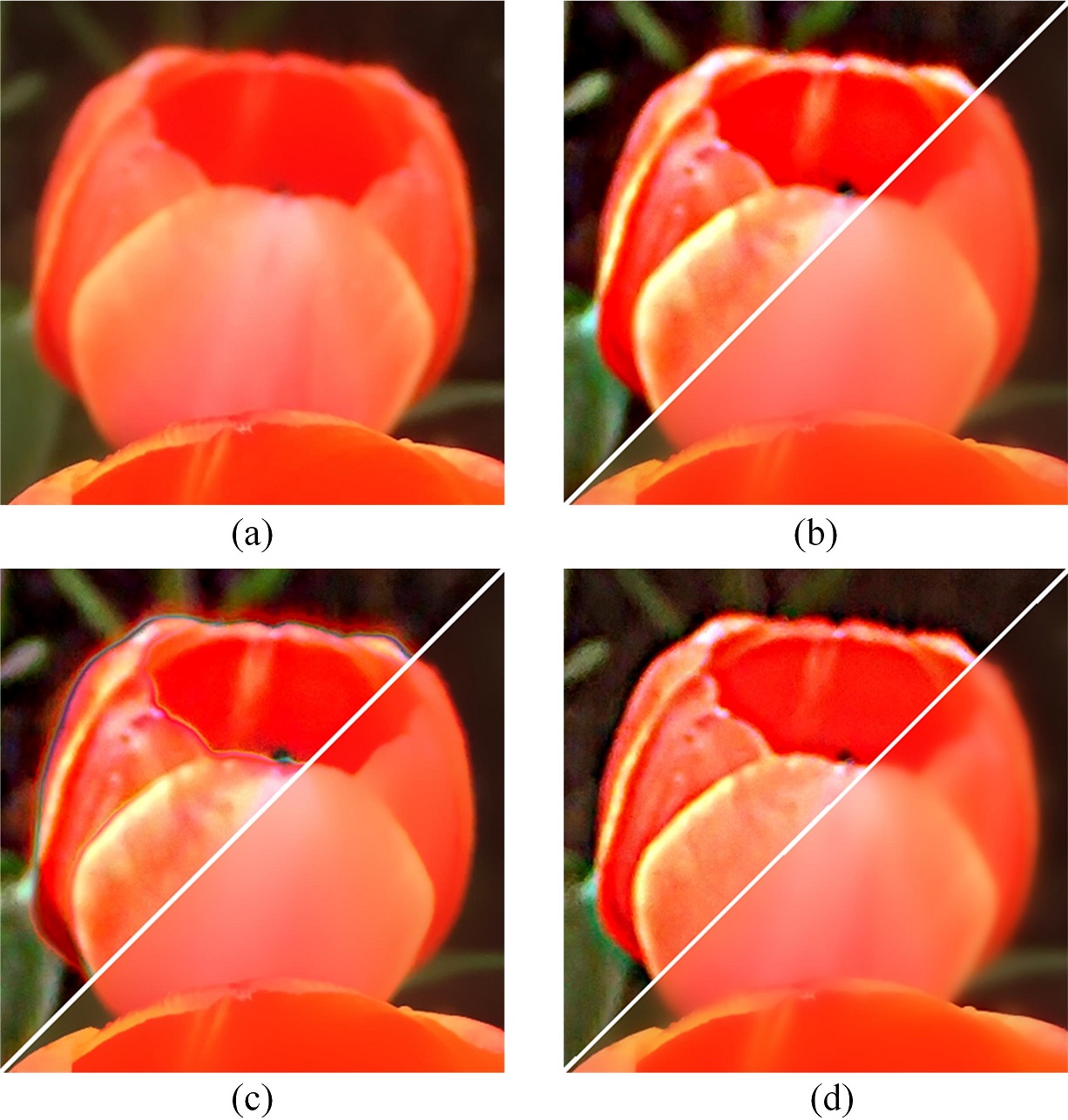}\\
  \caption{(a) Input image. Smoothed and detail enhanced images by (b) guided image filter \cite{he2013guided} with $r=16,\epsilon=0.1^2$, (c) bilateral filter \cite{tomasi1998bilateral} with $\sigma_s = 16, \sigma_r=0.1$ and (d) the proposed method applied to the bilateral filter with $\sigma_s = 16, \sigma_r = 0.025, \beta = 16$. Enhanced images are obtained by adding $5\times$ the detail layers to the input image in (a).}\label{FigCover}
\end{figure}

Edge-preserving image smoothing is a fundamental processing procedure in several low level computer vision applications, please refer to the applications in \cite{durand2002fast, farbman2010diffusion, farbman2008edge, gastal2011domain, gastal2012adaptive, hosni2013fast, petschnigg2004digital} for examples. For most smoothing filters in the literature, they assume the smoothed output images are piecewise constant. This can be simply modeled as:
\begin{equation}\label{EqPieceConstantModel}
    \hat{I}_\mathbf{p}=a_k, \ \mathbf{p}\in N(k)
\end{equation}
where $\hat{I}_\mathbf{p}$ represents the expected output pixel value at position $\mathbf{p}=[p_x, p_y]$, $a_k$ is the expected constant value of pixel values inside the $k$th region in the image which is denoted as $N(k)$. Note that $N(k)$ can have different definitions for different methods. For example, it can be a regular patch (e.g., square patch) in bilateral filter \cite{tomasi1998bilateral} while it can be an irregular image region separated by edges in gradient $L_0$ norm smoothing \cite{xu2011image}.

Filters of above assumption include bilateral filter \cite{tomasi1998bilateral}, joint bilateral filter \cite{petschnigg2004digital}, adaptive manifold filter \cite{gastal2012adaptive}, domain transform filter \cite{gastal2011domain}, median filter or weighted median filter \cite{zhang2014100+}, etc. The above mentioned filters are usually denoted as \emph{local methods} in the literature \cite{he2013guided}. There are also \emph{global methods} that are based on the piecewise constant model assumption in Eq.~(\ref{EqPieceConstantModel}) such as gradient $L_0$ norm smoothing \cite{xu2011image} and total variation smoothing \cite{rudin1992nonlinear}. In this paper, we denote the filters based on the piecewise constant model assumption in Eq.~(\ref{EqPieceConstantModel}) as piecewise constant filters. The corresponding filtering process is denoted as piecewise constant smoothing.

Although the piecewise constant model can meet the need of most applications, it may not work well for some special applications such as image detail enhancement, HDR tone mapping, etc. For these applications, a piecewise linear model is more preferred. Piecewise constant filters, especially those local methods mentioned above, can cause artifacts such as gradient reversals \cite{he2013guided} in these applications. Although most global methods are capable of handling with this kind of artifacts such as the weighted least squares framework in \cite{farbman2008edge}, there are also global methods that can cause such artifacts, for example, the gradient $L_0$ norm smoothing \cite{xu2011image} that will be shown in this paper.

In contrast to the piecewise constant model assumption, the recently proposed guided image filter \cite{he2013guided} and its variants \cite{dai2015fully, lu2012cross, tan2014multipoint} are based on a piecewise linear model which can be formulated as:
\begin{equation}\label{EqGFPieceLinearModel}
    \hat{I}_\mathbf{p}=a_kG_\mathbf{p} + b_k, \ \mathbf{p} \in \omega(k)
\end{equation}
where $\hat{I}_\mathbf{p}$ is the expected output pixel value at position $\mathbf{p}$, $G_\mathbf{p}$ is the corresponding pixel value of guidance image $G$ at position $\mathbf{p}$. $a_k$ and $b_k$ remain constant inside the window $\omega(k)$. The model in Eq.~(\ref{EqGFPieceLinearModel}) is piecewise linear with respect to the guidance image $G$. These methods show no gradient reversal artifacts in several applications where the bilateral filter \cite{tomasi1998bilateral} shows clear gradient reversal artifacts. We denote the guided filter \cite{he2013guided} and its variants \cite{dai2015fully, lu2012cross, tan2014multipoint} as piecewise linear filters and denote their corresponding smoothing procedures as piecewise linear smoothing.

In this paper, we also assume a piecewise linear model of the output image. However, our model is a spatially piecewise linear one which is different from the model of guided image filter \cite{he2013guided} in Eq.~(\ref{EqGFPieceLinearModel}). In addition, our method focuses on how to perform piecewise linear smoothing using classical piecewise constant filters. The contributions of this paper are as follows:
\begin{itemize}
  \item[1.] We show a simple yet very effective framework to perform piecewise linear smoothing built upon several classical piecewise constant filters including bilateral filter \cite{tomasi1998bilateral}, adaptive manifold filter \cite{gastal2012adaptive}, normalized convolution of domain transform filter \cite{gastal2011domain}, weighted median filter \cite{zhang2014100+} and gradient $L_0$ norm smoothing \cite{xu2011image}. Comprehensive experimental results of image detail enhancement, HDR tone mapping and flash/no-flash image filtering show that our method can perfectly eliminate gradient reversal artifacts caused by the original piecewise constant filters.
  \item[2.] Some accelerated methods such as the weighted median filter in \cite{zhang2014100+} need to quantize image intensity values into different bins. In most cases, a large number of bins are usually needed to avoid quantization artifacts. In contrast, we show that our method based on the weighted median filter \cite{zhang2014100+} can properly eliminate quantization artifacts with a much smaller number of bins, which greatly reduces the computational cost.
\end{itemize}

\section{The Method}

\subsection{Piecewise Linear Smoothing Using Piecewise Constant Filters}

Different from the piecewise linear model in Eq.~(\ref{EqGFPieceLinearModel}), we formulate images with a piecewise linear model as:
\begin{equation}\label{EqPieceLinearModel}
    \hat{I}_\mathbf{p}=\mathbf{a}_k^T\mathbf{p} + b_k, \ \mathbf{p} \in N(k)
\end{equation}
where $\mathbf{a}_k=[a_k^x, a_k^y]^T$ and $b_k$ are some linear coefficients that are assumed to remain constant inside $N(k)$. $N(k)$ is defined the same as that in Eq.~(\ref{EqPieceConstantModel}). In this way, Eq.~(\ref{EqPieceLinearModel}) is a spatially piecewise linear model. It is thus different from the model in Eq.~(\ref{EqGFPieceLinearModel}) which is linear with respect to the guidance image. In fact, Eq.~(\ref{EqPieceLinearModel}) is known as linear polynomial regression which is also adopted in \cite{buades2006staircasing} to eliminate the staircasing effect in image denoising. However, as we will show in the next paragraphs in this section, our method is different from the method in \cite{buades2006staircasing}. In their method, the linear coefficients (i.e., $\mathbf{a}_k$ and $b_k$) are explicitly calculated. In contrast, our method focuses on how to perform piecewise linear smoothing using piecewise constant filters where both $\mathbf{a}_k$ and $b_k$ do not need to be explicitly calculated.

Unlike the model in Eq.~(\ref{EqGFPieceLinearModel}) which is used for the explicit formulation of the filter \cite{he2013guided}, the model in Eq.~(\ref{EqPieceLinearModel}) is a general and abstract model which is not used for explicit formulations of filters in this paper. However, when we take the derivative of $\hat{I}_\mathbf{p}$ with respect to $\mathbf{p}$, then we have:
\begin{equation}\label{EqPieceLinearModelDerivative}
    \frac{\partial\hat{I}_\mathbf{p}}{\partial \mathbf{p}}=\mathbf{a}_k, \ \mathbf{p} \in N(k)
\end{equation}

\begin{figure*}
  \includegraphics[width=1\linewidth]{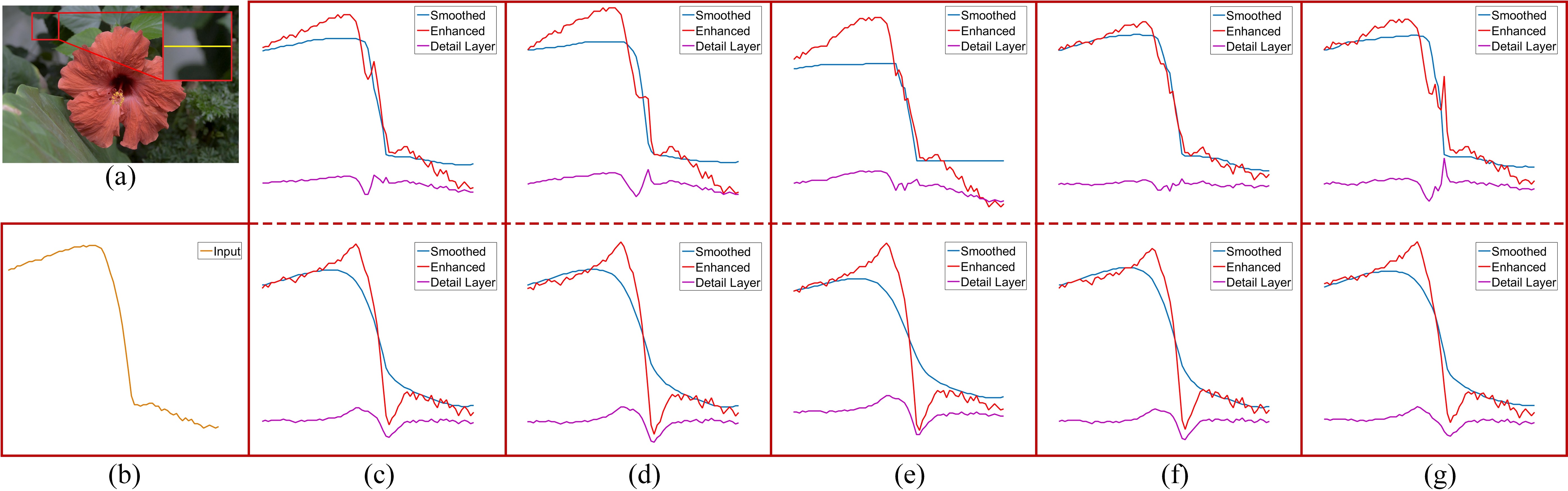}\\
  \caption{(b) 1D plot of the scan line in (a). The first row of (c)$\sim$(g) shows the results of classical piecewise constant filters. The results of the proposed piecewise linear smoothing applied to these piecewise constant filters are shown in the second row. Detail layers are obtained by subtracting the corresponding smoothed signals from the input shown in (b). Enhanced signals are obtained by adding $2\times$ the corresponding detail layers to the input shown in (b). (c) Adaptive manifold filter \cite{gastal2012adaptive} and (d) bilateral filter \cite{tomasi1998bilateral} with $\sigma_s=16, \sigma_r=0.1$ for piecewise constant smoothing and $\sigma_s=16, \sigma_r = 0.025, \beta=16$ for piecewise linear smoothing. (e) Gradient $L_0$ norm smoothing \cite{xu2011image} with $\lambda = 0.007$ for piecewise constant smoothing and $\lambda = 0.00175, \beta=16$ for piecewise linear smoothing. (f) Normalized convolution of domain transform filter \cite{gastal2011domain} with $\sigma_s=16, \sigma_r=0.1$ for piecewise constant smoothing and $\sigma_s=16, \sigma_r=0.025, \beta=16$ for piecewise linear smoothing. (g) Weighted median filter \cite{zhang2014100+} with $r=16, \sigma_r=0.1$ for piecewise constant smoothing and $r=16, \sigma_r=0.025, \beta=16$ for piecewise linear smoothing. Please refer to the corresponding papers for the detailed definition of the parameters in these piecewise constant filters.}\label{FigPCvsPL1D}
\end{figure*}

Note that $\frac{\partial\hat{I}_\mathbf{p}}{\partial \mathbf{p}}$ is by definition the gradient of the image at $\mathbf{p}$. Now it is simple to have the following conclusion: \emph{for images of the spatially piecewise linear model in Eq.~(\ref{EqPieceLinearModel}), their gradients are piecewise constant.} By comparing Eq.~(\ref{EqPieceLinearModelDerivative}) with Eq.~(\ref{EqPieceConstantModel}), it is clear that they are very similar. The only difference is that what on the left side of Eq.~(\ref{EqPieceConstantModel}) are image intensities while they are image gradients on the left side of Eq.~(\ref{EqPieceLinearModelDerivative}). Similarly, we can smooth image gradients with classical piecewise constant filters mentioned in Sec.~\ref{SecIntroduction}. However, the difference is that this procedure assumes the corresponding images of the smoothed output gradients are spatially piecewise linear as modeled in Eq.~(\ref{EqPieceLinearModel}).

However, the problem of this procedure is that we cannot simply reconstruct an image only from its smoothed gradients. Note that we also have the input image at the same time. The problem of using the input image and its filtered gradients to reconstruct the filtered image has already been studied by Xu et~al. \cite{xu2015deep}. For an input image $I_0$, its $x$-axes and $y$-axes gradients are denoted as $\partial_x I_0$ and $\partial_y I_0$ respectively. By denoting the smoothing process of piecewise constant filters as $\mathcal{F}_{pc}(\cdot)$, then the piecewise linear output image $\hat{I}$ can be reconstructed by minimizing the following energy proposed by Xu et~al. \cite{xu2015deep}:
\begin{equation}\label{EqReconstruct}
\small
    \|\hat{I}-I_0\|^2 + \beta\left\{\|\partial_x\hat{I} - \mathcal{F}_{pc}(\partial_xI_0)\|^2 + \|\partial_y\hat{I} - \mathcal{F}_{pc}(\partial_yI_0)\|^2\right\}
\end{equation}

Note that although we adopt the reconstruction procedure proposed by Xu et~al. \cite{xu2015deep}, our work is completely different from theirs. Their method focuses on implementing classical filters with deep neural networks. Their method still performs piecewise constant smoothing if their implementation is based on a piecewise constant filter. In contrast, our method focuses on how to perform piecewise linear smoothing using piecewise constant filters.

Based on the statement above, we can perform piecewise linear smoothing using piecewise constant filters in the following two steps: (1) Smoothing the $x$-axes and $y$-axes gradients $\partial_x I_0$ and $\partial_y I_0$ of the input image $I_0$ with $\mathcal{F}_{pc}(\cdot)$. $\mathcal{F}_{pc}(\cdot)$ can be any piecewise constant filter such as bilateral filter \cite{tomasi1998bilateral} and gradient $L_0$ norm smoothing \cite{xu2011image}. The smoothed output gradients are denoted as $\mathcal{F}_{pc}(\partial_xI_0)$ and $\mathcal{F}_{pc}(\partial_yI_0)$. (2) Using Eq.~(\ref{EqReconstruct}) to reconstruct the output image $\hat{I}$ from $I_0$, $\mathcal{F}_{pc}(\partial_xI_0)$ and $\mathcal{F}_{pc}(\partial_yI_0)$ with a proper value of $\beta$. Then the reconstructed $\hat{I}$ is spatially piecewise linear as modeled in Eq.~(\ref{EqPieceLinearModel}).

\begin{figure*}
  \includegraphics[width=1\linewidth]{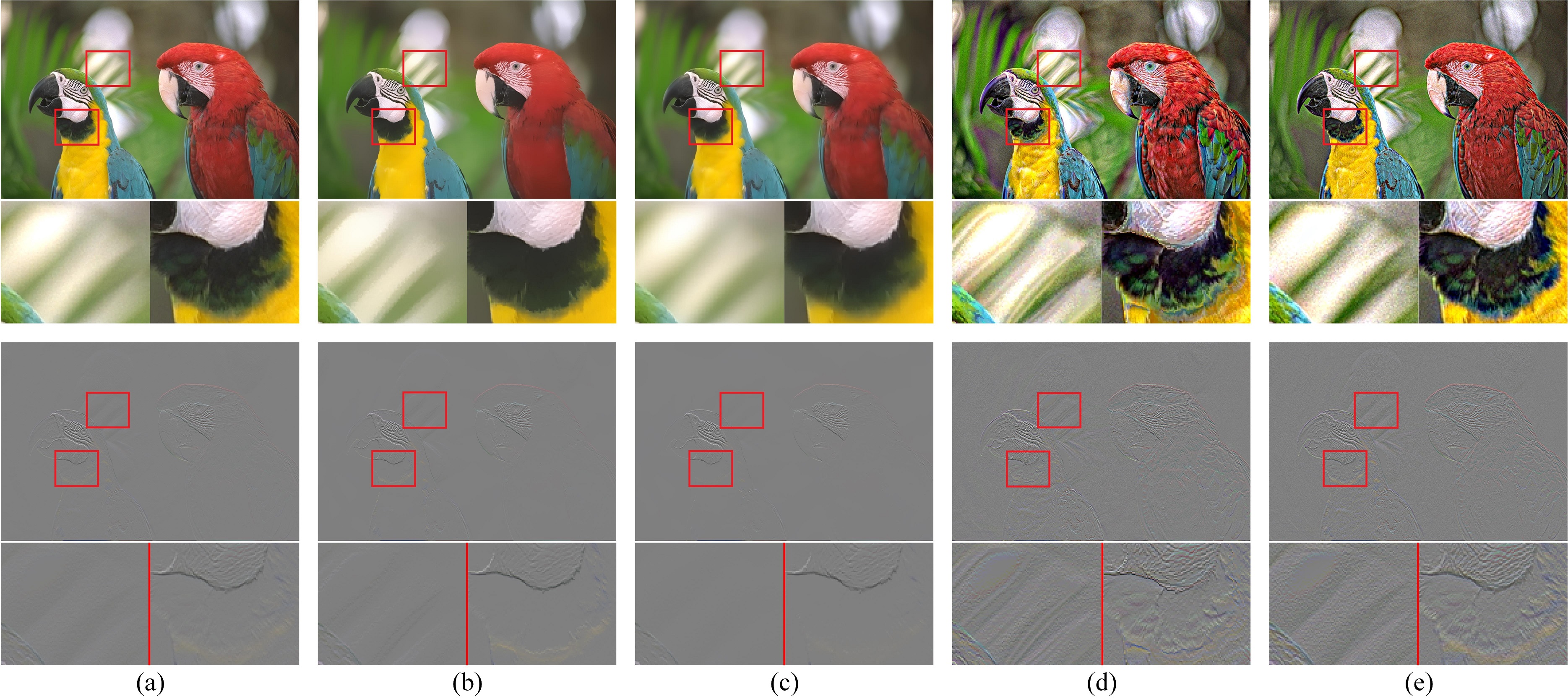}\\
  \caption{(a) Input image. (b) Smoothed image and (d) detail enhanced image by adaptive manifold filter \cite{gastal2012adaptive} with $\sigma_s=16,\sigma_r=0.1$. (c) Smoothed image and (e) detail enhanced image by the proposed method applied to adaptive manifold filter \cite{gastal2012adaptive} with $\sigma_s = 16, \sigma_r = 0.025, \beta = 16$. Enhanced images are obtained by adding $5\times$ the detail layers to the input image in (a). The corresponding $y$-axes gradient images are shown in the second row. Please pay attention to the highlighted regions in (b) where weak edges are improperly sharpened by adaptive manifold filter \cite{gastal2012adaptive}. This results in the gradient reversal artifacts in (d). Best viewed on screen.}\label{FigPCvsPL}
\end{figure*}

\begin{figure*}
  \includegraphics[width=1\linewidth]{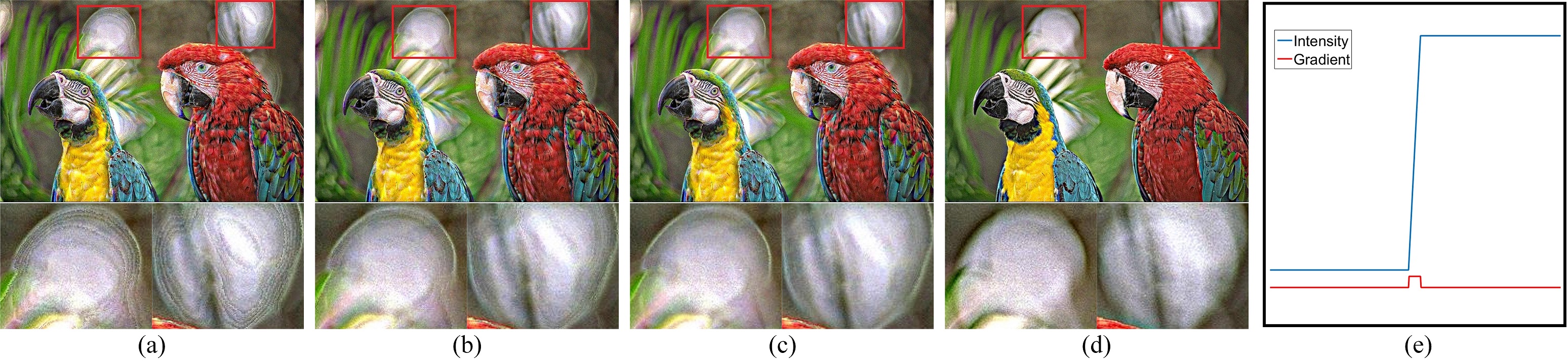}\\
  \caption{Image detail enhancement with weighted median filter \cite{zhang2014100+} that quantizes image intensity values into (a) $2^8$ bins where clear quantization artifacts exist. Time cost is 1.23 seconds. (b) $2^{10}$ bins where weak quantization artifacts exist. Time cost is 3.15 seconds. (c) $2^{12}$ bins where no quantization artifacts exist. Time cost is 19.32 seconds. (d) Our method applied to weighted median filter \cite{zhang2014100+} with gradient values quantized into $2^8$ bins. No quantization artifacts exist. Time cost is 2.56 seconds. (e) Image intensities of very large value can have gradients of quite small amplitude. Best viewed on screen.}\label{FigTimeReduce}
\end{figure*}

\begin{figure*}
  \includegraphics[width=1\linewidth]{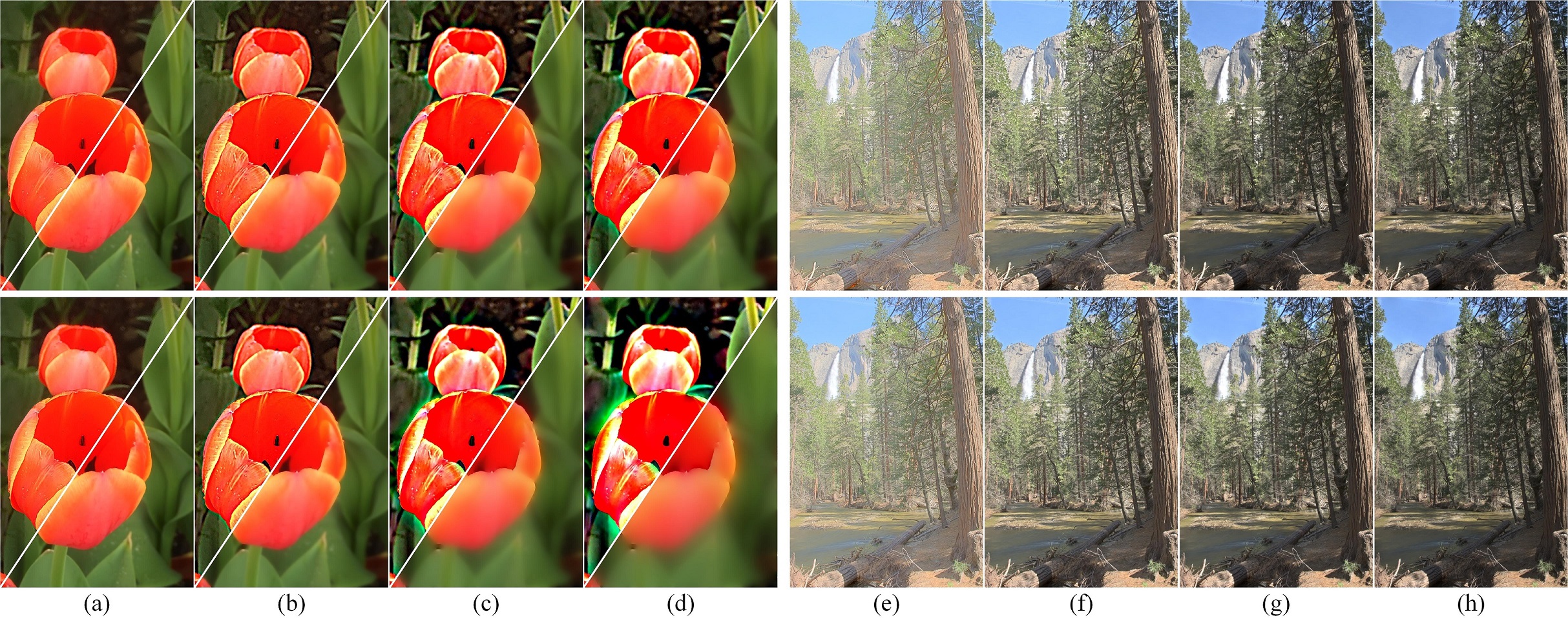}\\
  \vspace{-0.5em}\caption{(a)$\sim$(d) Image detail enhancement and (e)$\sim$(h) HDR tone mapping with different values of $\beta$ in Eq.~(\ref{EqReconstruct}). The first row shows results of the proposed piecewise linear smoothing using adaptive manifold filter \cite{gastal2012adaptive} with $\sigma_s=16, \sigma_r=0.025$ for image detail enhancement and $\sigma_s=16, \sigma_r=0.03$ for HDR tone mapping. The second row shows results of the proposed piecewise linear smoothing using gradient $L_0$ norm smoothing \cite{xu2011image} with $\lambda=0.00175$ for image detail enhancement and $\lambda = 0.0175$ for HDR tone mapping. Enhanced images are obtained by adding $5\times$ the corresponding detail layers to the input image. (a) $\beta=1$. (b) $\beta=16$. (c) $\beta=256$. (d) $\beta=1024$. (e) $\beta=16$. (f) $\beta=64$. (g) $\beta=256$. (h) $\beta=1024$.}\label{FigDifferentBeta}\vspace{-1em}
\end{figure*}

\subsection{Piecewise Constant Smoothing vs Piecewise Linear Smoothing}

Firstly, we compare the smoothing quality of classical piecewise constant smoothing and the proposed piecewise linear smoothing applied to these piecewise constant filters. As stated in Sec.~\ref{SecIntroduction}, classical piecewise constant filters are more suitable for piecewise constant images regions. When the image region is more likely to be piecewise linear, piecewise constant filters may cause artifacts. We show a 1D illustration in Fig.~\ref{FigPCvsPL1D}(b) where a piecewise linear model is more suitable for this 1D signal. We show the smoothed signals, their corresponding detail layers and enhanced signals by classical piecewise constant filters in the first row of Fig.~\ref{FigPCvsPL1D}(c)$\sim$(g). Clearly, the detail layers cannot correctly reflect the details in the original signal. As a result, clear artifacts (known as gradient reversals) exist in the enhanced signals. In contrast, all the detail layers obtained by the proposed piecewise linear smoothing in the second row of Fig.~\ref{FigPCvsPL1D}(c)$\sim$(g) can properly reflect the details in the original signal. Thus, no gradient reversal artifacts exist in their enhanced signals.

To further compare the behavior of piecewise constant smoothing and piecewise linear smoothing for strong edges and weak edges, we show a more clear visual comparison in the gradient domain in Fig.~\ref{FigPCvsPL}. As shown in highlighted regions, both two kinds of smoothing can properly preserve strong gradients which means strong edges are properly preserved. For weak edges of small gradients which should be smoothed out, piecewise linear smoothing can properly smooth out these small gradients as shown in the second row of Fig.~\ref{FigPCvsPL}(c). However, these gradients are not properly smoothed by piecewise constant smoothing or even improperly sharpened as shown in the second row of Fig.~\ref{FigPCvsPL}(b). As a result, clear gradient reversal artifacts exist in its enhanced image and the corresponding gradients in Fig.~\ref{FigPCvsPL}(d) while no gradient reversal artifacts exist in the enhanced results of piecewise linear smoothing in Fig.~\ref{FigPCvsPL}(e). This gradient domain comparison shows that the proposed piecewise linear smoothing is more accurate than classical piecewise constant smoothing in term of ``edge-preserving smoothing'' which means to smooth weak edges but preserve strong edges.

Secondly, we compare the computational cost of classical piecewise constant smoothing and their corresponding piecewise linear smoothing. There are two steps in the proposed piecewise linear smoothing. The first step needs to smooth both $x$-axes gradients and $y$-axes gradients with classical piecewise constant filters. The computational cost of this step is twice of that of the corresponding classical piecewise constant smoothing. The second step requires to solve a linear system of which the computational cost only depends on the image size. We use preconditioned conjugate gradient (PCG) to speed up the sparse linear system with the incomplete Cholesky factorization preconditioner \cite{szeliski2006locally} which is the same as that in \cite{xu2015deep}. This step costs $\sim$0.3 seconds for a one-megapixel RGB image. In summary, the computational cost of the proposed piecewise linear smoothing is $\sim$0.3 seconds plus twice of that of the corresponding piecewise constant smoothing for a one-megapixel RGB image.

\begin{figure}
  \includegraphics[width=1\linewidth]{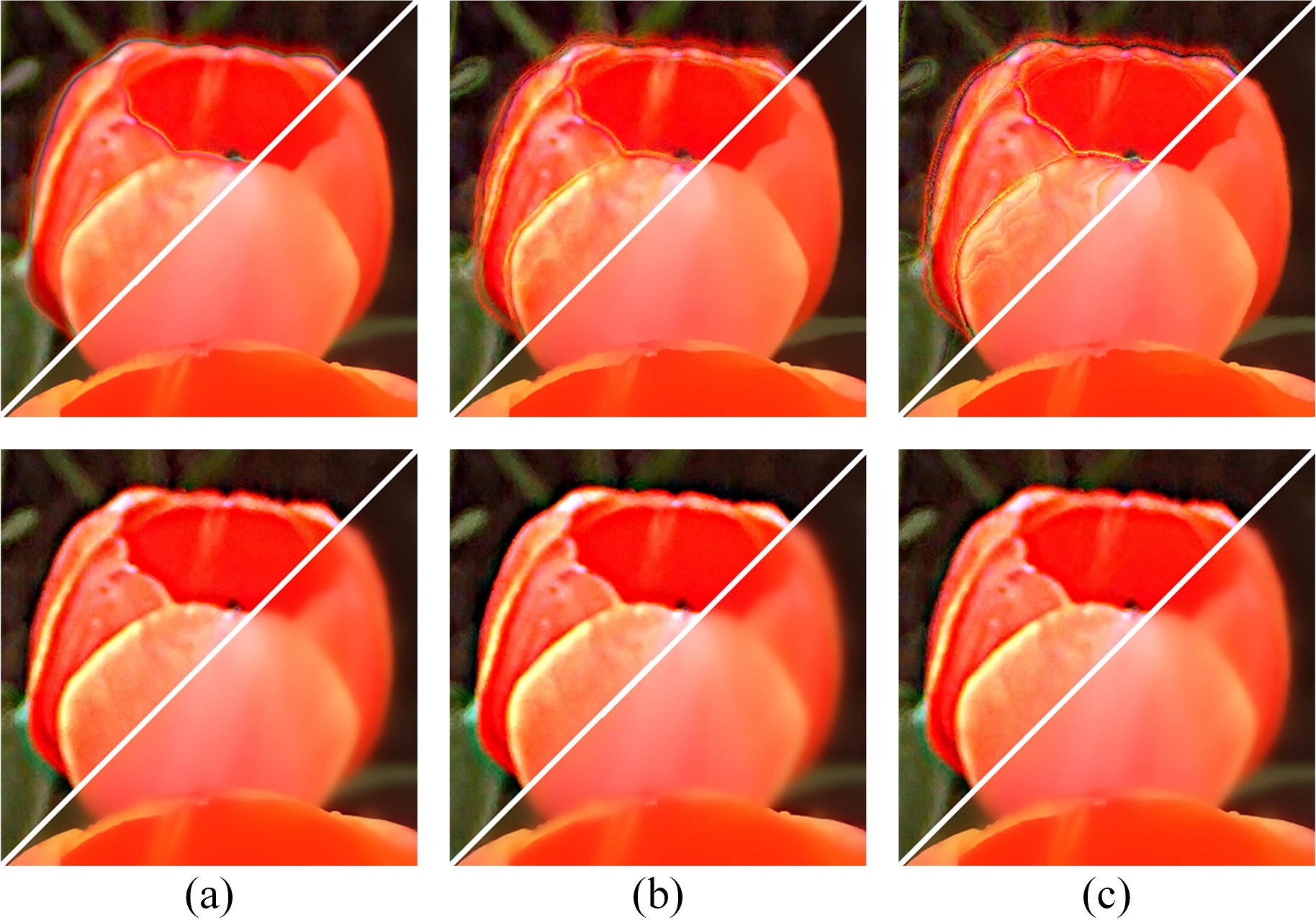}\\
  \vspace{-0.5em}\caption{Image detail enhancement with different strategies. Smoothed and enhanced images in the first row are obtained by firstly smoothing the input image with piecewise constant filters and then using the gradients of the smoothed image to reconstruct the final smoothed image using Eq.~(\ref{EqReconstruct}). Parameters for piecewise constant smoothing are the same as those in Fig.~(\ref{FigPCvsPL1D}) with $\beta=16$. Smoothed and enhanced images in the second row are obtained by the proposed piecewise linear smoothing. Parameters are the same as those of the piecewise linear smoothing in Fig.~(\ref{FigPCvsPL1D}) with $\beta=16$. (a) Bilateral filter \cite{tomasi1998bilateral}. (b) Gradient $L_0$ norm smoothing \cite{xu2011image}. (c) Weighted median filter \cite{zhang2014100+}.}\label{FigDiffGradReconstruction}\vspace{-1em}
\end{figure}

However, for accelerated methods such as the weighted median filter in \cite{zhang2014100+} which need to quantize image intensity values into different bins, we find our method can further reduce the computational cost which is in contrast to the analysis above. For most cases, the accelerated piecewise constant filters usually need a large number of bins to avoid quantization artifacts while the proposed piecewise linear smoothing only needs a much smaller number of bins to achieve similar results. Fig.~\ref{FigTimeReduce}(a) shows image detail enhancement with the weighted median filter in \cite{zhang2014100+} using $2^8$ quantization bins where clear quantization artifacts exist. When adopting $2^{10}$ bins, weak quantization artifacts still exist as shown in highlighted regions in Fig.~\ref{FigTimeReduce}(b). These quantization artifacts disappear when adopting $2^{12}$ bins as shown in Fig.~\ref{FigTimeReduce}(c). In contrast, the result of the proposed piecewise linear smoothing in Fig.~\ref{FigTimeReduce}(d) is obtained with only $2^8$ quantization bins and no quantization artifacts exist. This greatly reduces the computational cost from 19.32 seconds to 2.56 seconds. The reason for this phenomena is clear. The classical piecewise constant smoothing is performed in the intensity domain where intensity values could be very large as illustrated in Fig.~\ref{FigTimeReduce}(e). In contrast, the first step of the proposed piecewise linear smoothing is performed in the gradient domain. Very large intensity values can always have gradients of very small amplitude as illustrated in Fig.~\ref{FigTimeReduce}(e). Thus, to achieve a similar quantization error precision, gradient domain only needs much fewer quantization bins than intensity domain does.

\subsection{Further Discussions}

\begin{figure*}
  \includegraphics[width=1\linewidth]{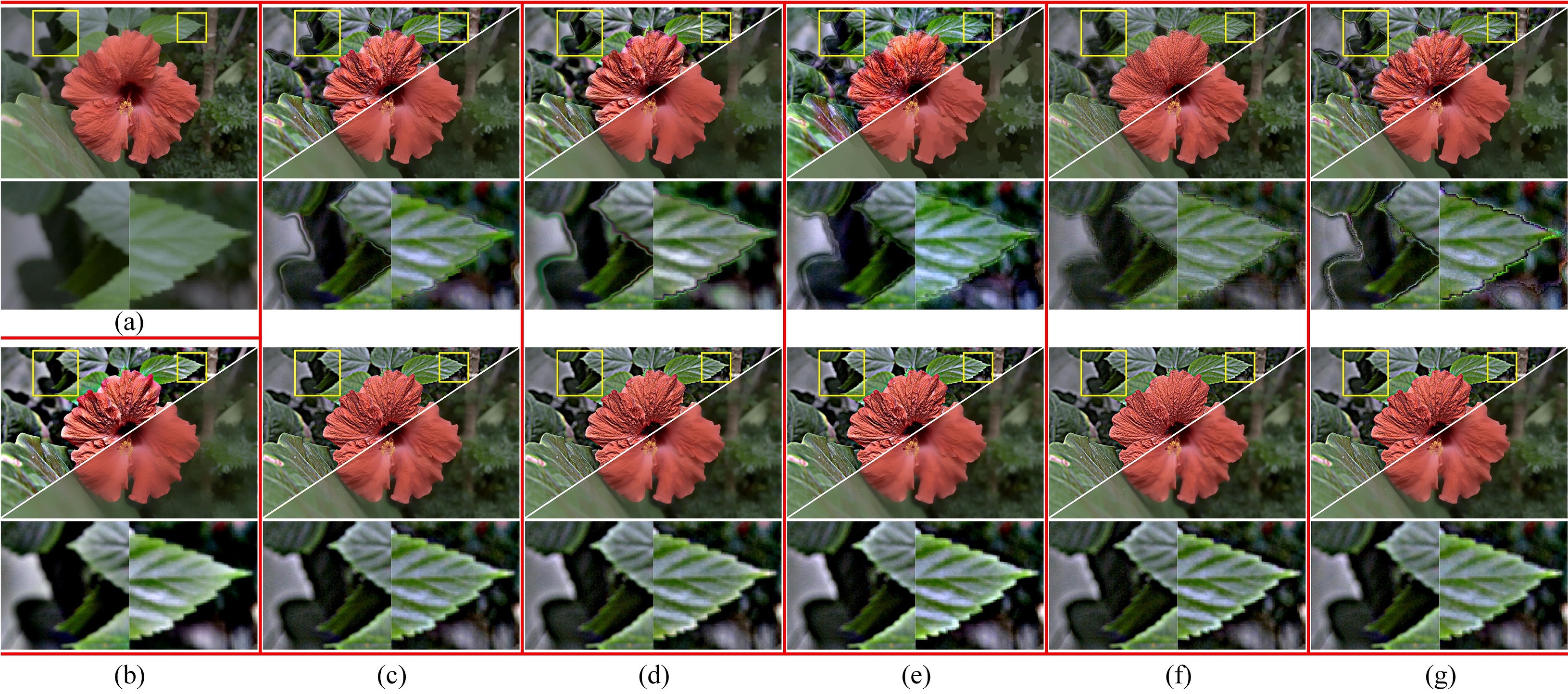}\\
  \caption{(a) Input image. (b) Smoothed and detail enhanced image by guided image filter \cite{he2013guided} with $r=16, \epsilon=0.1^2$. The first row of (c)$\sim$(g) shows smoothed and enhanced images by classical piecewise constant filters. The results of the proposed piecewise linear smoothing applied to these filters are shown in the second row. Enhanced images are obtained by adding $5\times$ the corresponding detail layers to the input shown in (a). (c) Adaptive manifold filter \cite{gastal2012adaptive}. (d) Bilateral filter \cite{tomasi1998bilateral}. (e) Gradient $L_0$ norm smoothing \cite{xu2011image}. (f) Normalized convolution of domain transform filter \cite{gastal2011domain}. (g) Weighted median filter \cite{zhang2014100+}. Parameters of the filters in (c)$\sim$(g) are the same as those in Fig.~(\ref{FigPCvsPL1D}).}\label{FigDetailEnhancement}\vspace{-1em}
\end{figure*}

Compared with classical piecewise constant smoothing, there is one additional parameter of the proposed piecewise linear smoothing which is $\beta$ in Eq.~(\ref{EqReconstruct}). A natural question is how to choose the value of $\beta$ for different filters and applications. In the work of Xu et~al. \cite{xu2011image}, the value of $\beta$ is decided as the one that has highest PSNR between the output and the ``ground truth'' which is the output of classical piecewise constant filters. However, there is no ``ground truth'' for the proposed piecewise linear smoothing. Thus, the value of $\beta$ is empirically decided for different filters and applications. Fig.~\ref{FigDifferentBeta} shows example results obtained with different values of $\beta$. A clear observation is that increasing the value of $\beta$ can increase the smoothing strength. In our experiments, we empirically set $\beta=16$ for all the filters in image detail enhancement and $\beta=64$ for all the filters in HDR tone mapping. For flash/no-flash filtering, we empirically set $\beta=128$ to better smooth the heavy noise in no-flash images. Note that other parameter settings may also achieve promising results.

As stated above, $\beta$ can affect the final smoothing results which means that Eq.~(\ref{EqReconstruct}) can also affect the final smoothing results. A rigorous question is that whether the improvement of the proposed piecewise linear smoothing over the classical piecewise constant smoothing results from the image reconstruction step of Eq.~(\ref{EqReconstruct}). To answer this question, we perform another experiment: using the gradients of images smoothed by classical piecewise constant filters to reconstruct the final smoothed images. Note that this procedure is to perform piecewise constant smoothing. We show comparison with the proposed piecewise linear smoothing in Fig.~\ref{FigDiffGradReconstruction}. Clearly, using gradients of images smoothed by classical piecewise constant filters for image reconstruction still has the problem of gradient reversal artifacts. Thus, we can conclude that the improvement of the proposed piecewise linear smoothing results from the spatially piecewise linear model in Eq.~(\ref{EqPieceLinearModel}).

\section{Applications and Experimental Results}

We apply our method to several classical piecewise constant filters to perform piecewise linear smoothing. These filters include bilateral filter \cite{tomasi1998bilateral}, adaptive manifold filter \cite{gastal2012adaptive}, normalized convolution of domain transform filter \cite{gastal2011domain}, gradient $L_0$ norm smoothing \cite{xu2011image} and weighted median filter \cite{zhang2014100+}. For bilateral filter, we adopt its fast implementation in \cite{paris2006fast}\footnote{Source code can be downloaded here: \url{http://people.csail.mit.edu/jiawen/software/bilateralFilter.m}}. Implementation of the other filters can be downloaded from their project websites. We perform experiments in image detail enhancement, HDR tone mapping and flash/no-flash image filtering to validate the effectiveness of the proposed method. In all applications, input images, including the gradients in the proposed piecewise linear smoothing, are firstly normalized into range $[0, 1]$ before smoothing and then normalized back to their original range after smoothing. For the problem of parameter setting, we adopt the ones used in previous papers. For example, the parameters of bilateral filter \cite{tomasi1998bilateral} for all the applications in this paper are the same as those adopted in the paper of guide image filter \cite{he2013guided}. For adaptive manifold \cite{gastal2012adaptive}, domain transform filter \cite{gastal2011domain} and weighted median filter, we set their parameters analogous to those of bilateral filter for fair comparisons. The parameter of gradient $L_0$ norm smoothing \cite{xu2011image} is set the same as that in its original paper. \textbf{For more experimental results, please refer to our supplementary materials.}

\begin{figure*}
  \includegraphics[width=1\linewidth]{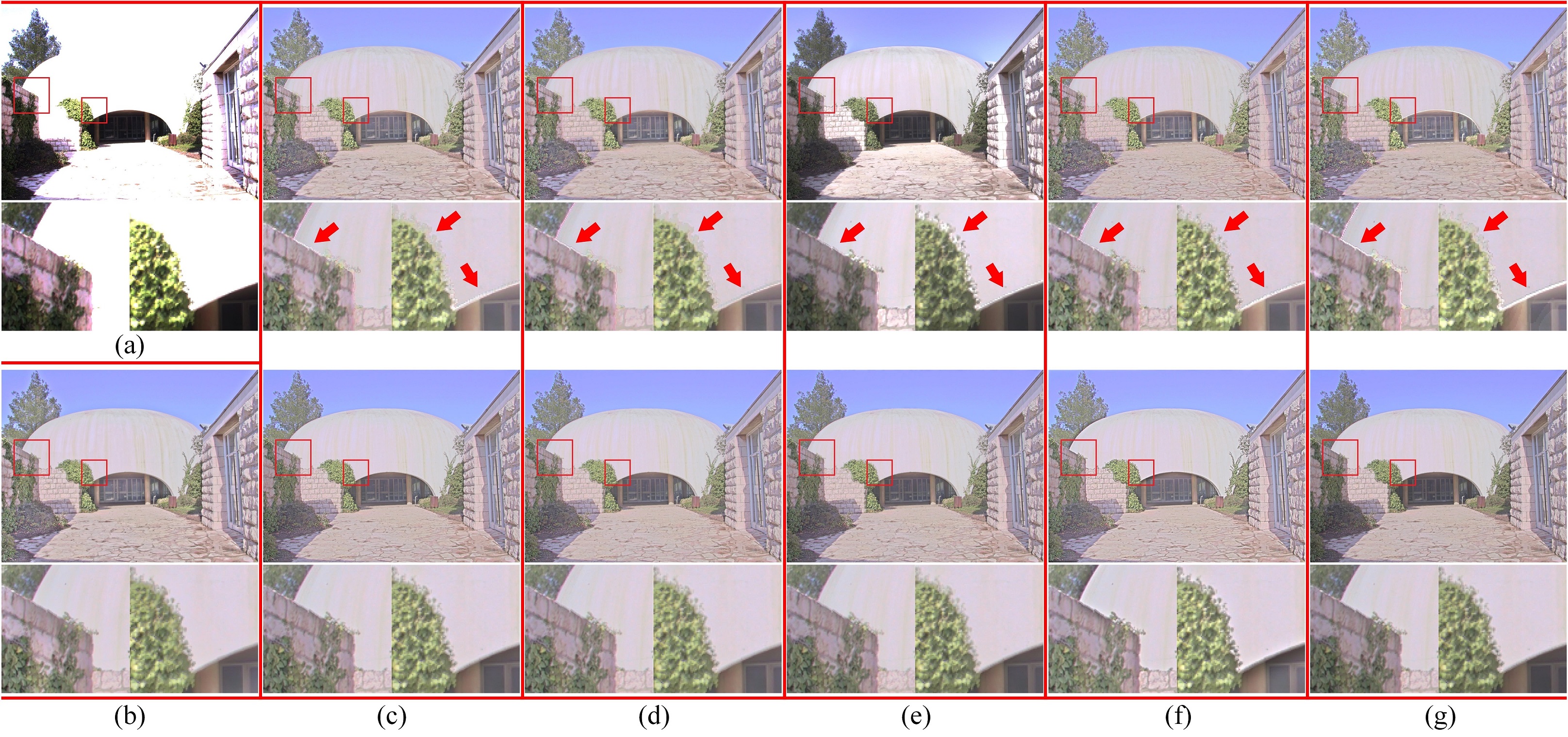}\\
  \caption{(a) Input HDR image. (b) HDR tone mapping result of guided image filter \cite{he2013guided} with $r=16, \epsilon=0.12^2$. The first row of (c)$\sim$(g) shows HDR tone mapping results of classical piecewise constant filters. The results of the proposed piecewise linear smoothing applied to these filters are shown in the second row. (c) Adaptive manifold filter \cite{gastal2012adaptive} and (d) bilateral filter \cite{tomasi1998bilateral} with $\sigma_s=16, \sigma_r=0.12$ for piecewise constant smoothing and $\sigma_s=16, \sigma_r = 0.03, \beta=64$ for piecewise linear smoothing. (e) Gradient $L_0$ norm smoothing \cite{xu2011image} with $\lambda = 0.07$ for piecewise constant smoothing and $\lambda = 0.0175, \beta=64$ for piecewise linear smoothing. (f) Normalized convolution of domain transform filter \cite{gastal2011domain} with $\sigma_s=16, \sigma_r=0.12$ for piecewise constant smoothing and $\sigma_s=16, \sigma_r=0.03, \beta=64$ for piecewise linear smoothing. (g) Weighted median filter \cite{zhang2014100+} with $r=16, \sigma_r=0.12$ for piecewise constant smoothing and $r=16, \sigma_r=0.03, \beta=64$ for piecewise linear smoothing. Best viewed on screen.}\label{FigHDRToneMapping}
\end{figure*}

\begin{figure*}
  \includegraphics[width=1\linewidth]{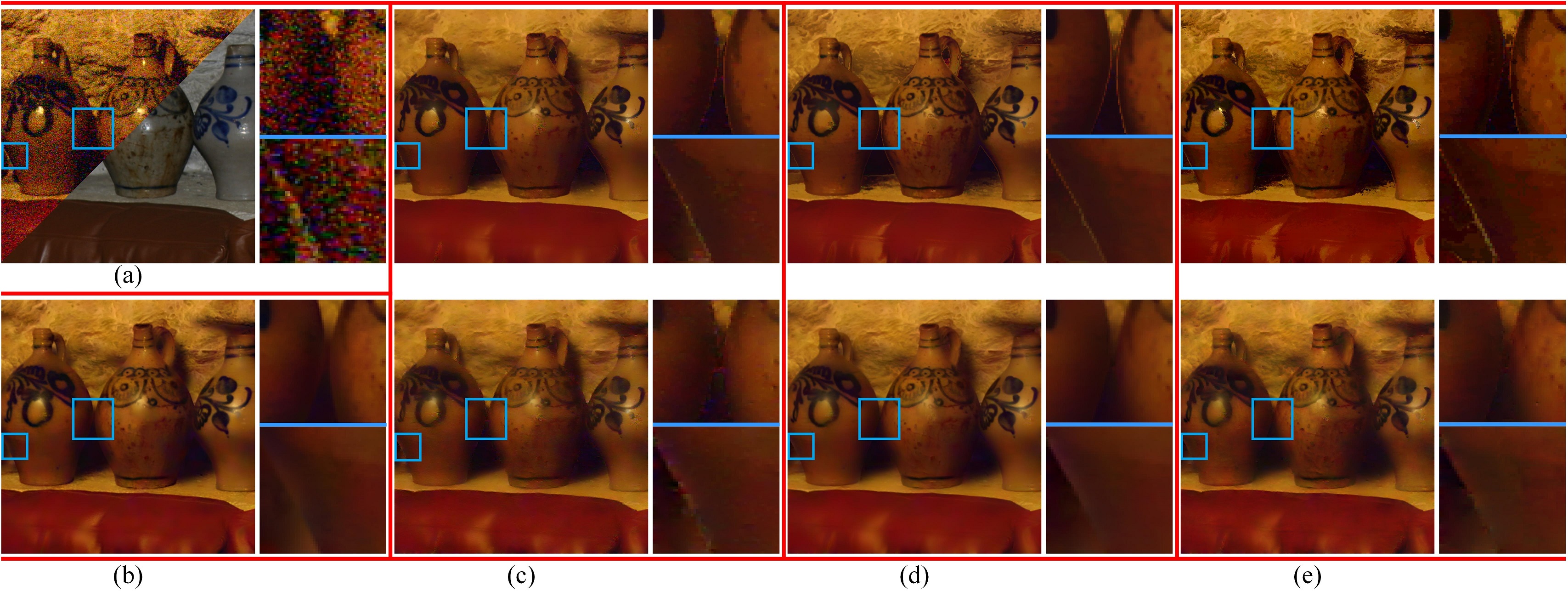}\\
  \caption{(a) Input flash image and no-flash image. (b) Smoothed no-flash image by guided image filter \cite{he2013guided} with $r=16, \epsilon=0.02^2$. The first row of (c)$\sim$(e) shows smoothed no-flash images by classical piecewise constant filters. The results of the proposed piecewise linear smoothing applied to these filters are shown in the second row. (c) Adaptive manifold filter \cite{gastal2012adaptive} with $\sigma_s=16, \sigma_r=0.06$ for piecewise constant smoothing and $\sigma_s=16, \sigma_r = 0.015, \beta=128$ for piecewise linear smoothing. (d) Bilateral filter \cite{tomasi1998bilateral} with $\sigma_s=16, \sigma_r=0.02$ for piecewise constant smoothing and $\sigma_s=16, \sigma_r = 0.005, \beta=128$ for piecewise linear smoothing. (e) Weighted median filter \cite{zhang2014100+} with $r=16, \sigma_r=0.02$ for piecewise constant smoothing and $r=16, \sigma_r=0.005, \beta=128$ for piecewise linear smoothing. Best viewed on screen. }\label{FigFlashNoFlash}
\end{figure*}

\textbf{Image detail enhancement} is a fundamental application of image smoothing. By subtracting the smoothed image from the original input image, we obtain the detail layer which contains the high-frequency details of the original input image. Then the detail layer is magnified and added back to the original input image to get the detail enhanced image. If edges in the smoothed image are sharper than those in the original input image, then these edges will be boosted in a reverse direction which causes the gradient reversal artifacts \cite{farbman2008edge, he2013guided}. Gradient reversal artifacts are quite common in local piecewise constant filters such as the ones mentioned in the first paragraph of this section. There are also global methods such as the gradient $L_0$ norm smoothing \cite{xu2011image} that also have such problems. Illustrations in both Fig.~\ref{FigPCvsPL1D} and Fig.~\ref{FigPCvsPL} clearly show that images smoothed by piecewise constant filters can improperly sharpen edges which causes gradient reversal artifacts. In contrast, when the proposed method is applied to these piecewise constant filters to perform piecewise linear smoothing, all the edges are properly smoothed which can also be observed in Fig.~\ref{FigPCvsPL1D} and Fig.~\ref{FigPCvsPL}. Thus no gradient reversal artifacts exist in the enhanced images. Results in Fig.~\ref{FigDetailEnhancement} correspond to the 1D illustration in Fig.~\ref{FigPCvsPL1D}. In addition, we also show the result of guided image filter \cite{he2013guided} in Fig.~\ref{FigDetailEnhancement}(b) as a reference. The proposed piecewise linear smoothing can achieve artifacts free results which are similar to the result of guided image filter \cite{he2013guided}.

\textbf{HDR tone mapping} is another popular application that needs edge-preserving smoothing. It can be achieved by decomposing an HDR image into a piecewise smooth base layer conveying most of the energy and a detail layer \cite{durand2002fast, farbman2008edge, li2005compressing}. The base layer is the smoothed output of input HDR image. The base layer is then nonlinearly mapped to a low dynamic range and is re-combined with the detail layer to get the final tone mapped image. We adopt the framework in \cite{durand2002fast} where layer decomposition is applied to the logarithmic HDR images. Similar to image detail enhancement, tone mapped images using piecewise constant filters may also have gradient reversal artifacts. We show comparison between results of classical piecewise constant smoothing and the proposed piecewise linear smoothing in Fig.~\ref{FigHDRToneMapping}. Result of guided image filter \cite{he2013guided} is show in Fig.~\ref{FigHDRToneMapping}(b) as a reference of piecewise linear smoothing. Clear gradient reversal artifacts exist in results of classical piecewise constant smoothing in the first row of Fig.~\ref{FigHDRToneMapping} as shown in highlighted regions. In addition, small structures such as the leaves in highlighted regions also vanished in the tone mapped image. However, all these artifacts are properly eliminated in the results of the proposed piecewise linear smoothing in the second row of Fig.~\ref{FigHDRToneMapping}.

\textbf{Flash/no-flash filtering} was proposed by Petschnigg et~al. \cite{petschnigg2004digital} to smooth a noisy no-flash image with the guidance of the flash image. In \cite{he2013guided}, He et~al. showed that gradient reversal artifacts also existed in the smoothed results of joint bilateral filter \cite{petschnigg2004digital}. In our experiments, we find that results of adaptive manifold filter \cite{gastal2012adaptive} and weighted median filter \cite{zhang2014100+} also suffer from this problem. We show their results in the first row of Fig.~\ref{FigFlashNoFlash} and illustrate gradient reversals in highlighted regions. However, these artifacts are properly eliminated by the proposed piecewise linear smoothing applied to these piecewise constant filters, which is similar to the result of guided image filter \cite{he2013guided} in Fig.~\ref{FigFlashNoFlash}(b). When smoothing the gradients of no-flash images in the proposed method, we use the gradients of flash images as guidance images.

\section{Conclusion}

In this paper, we proposed a general framework to perform piecewise linear smoothing using several classical piecewise constant filters. The smoothed output images of the proposed method are based on a spatially piecewise linear model assumption which is different from the piecewise constant model assumption in most classical piecewise constant filters. The proposed method can properly handle with the problem of gradient reversal artifacts caused by the piecewise constant model assumption in several applications. In addition, the proposed method can further reduce the computational cost of accelerated methods which need to quantize image intensity values into different bins. Comprehensive experimental results in various applications show the effectiveness of the proposed method.

{\small
\bibliographystyle{ieee}
\bibliography{egbib}
}

\end{document}